\documentclass{iopart}
\usepackage{iopams}
\usepackage{setstack}
\usepackage{epsf}
\usepackage{graphics}

\begin{document}

\title[Aperiodic model for interacting polymers]
{Critical properties of an aperiodic model for interacting polymers}

\author{T A S Haddad\dag, R F S Andrade\ddag\ and S R Salinas\dag}
\address{\dag\ Instituto de F\'{\i}sica, Universidade de S\~{a}o Paulo, 
Caixa Postal 66318, 05315-970, S\~{a}o Paulo, SP, Brazil}
\address{\ddag\ Instituto de F\'{\i}sica, Universidade Federal da Bahia, 
Campus da Federa\c{c}\~{a}o, 40210-340, Salvador, BA, Brazil}

\begin{abstract}
We investigate the effects of aperiodic interactions on the critical
behavior of an interacting two-polymer model on hierarchical lattices 
(equivalent to the Migadal-Kadanoff approximation for the model on Bravais 
lattices), via renormalization-group and tranfer-matrix calculations. 
The exact renormalization-group recursion relations always present a 
symmetric fixed point, associated with the critical behavior of the 
underlying uniform model. If the aperiodic interactions, defined by s
ubstitution rules, lead to relevant geometric fluctuations, this fixed
point becomes fully unstable, giving rise to novel attractors of different 
nature. We present an explicit example in which this new attractor is a 
two-cycle, with critical indices different from the uniform model. In case of
the four-letter Rudin-Shapiro substitution rule, we find a surprising
closed curve whose points are attractors of period two, associated
with a marginal operator. Nevertheless, a scaling analysis indicates
that this attractor may lead to a new critical universality class.
In order to provide an independent confirmation of the scaling results,
we turn to a direct thermodynamic calculation of the specific-heat
exponent. The thermodynamic free energy is obtained from a transfer
matrix formalism, which had been previously introduced for spin systems,
and is now extended to the two-polymer model with aperiodic interactions. 
\end{abstract}

\pacs{64.60.Ak, 64.60.Cn, 61.44.-n}

\submitto{\JPA}

\ead{thaddad@if.usp.br}

\maketitle

\section{Introduction}

In some recent publications \cite{haddad1}-\cite{andrade2}, we have
used renormalization-group (RG) and transfer-matrix (TM) techniques
to investigate the effects of aperiodically distributed (but not disordered)
interactions on the critical behavior of ferromagnetic spin models.
In a real-space renormalization calculation for simple hierarchical
structures, we have written exact recursion relations in order to
show that relevant geometric (aperiodic) fluctuations play a very
similar role to disorder. These calculations lead to the formulation
of an exact extension for deterministic, aperiodic interactions of
the well-known Harris criterion for the relevance of disorder \cite{harris},
in a form coincident with Luck's heuristic, general derivation of
this extension \cite{luck}. Also, we have shown that relevant geometric
fluctuations give rise to distinct critical exponents, associated
with the appearance of new attractors in parameter space. The independent
transfer-matrix calculations have confirmed these results and provided
deeper insight on more refined details of the thermodynamics of aperiodic
spin systems, such as log-periodic oscillations of thermodynamic functions.

Now we consider a model of two directed polymers on a diamond hierarchical
lattice, with an aperiodic, layered distribution of interaction energies,
according to various substitution rules (see \cite{grimm,queffelec}
for extensive reviews of substitution sequences and their applications
to statistical models). Although the qualitative description of the
critical behavior is essentially similar, the exact renormalization-group
recursion relations in parameter space turn out to be much simpler
as compared to the calculations for spin systems. The case of two-letter
substitution rules is particularly simple. For irrelevant geometric
fluctuations, the critical behavior is governed by a symmetric fixed
point, with no changes with respect to the uniform case. For relevant
fluctuations, we show that this symmetric fixed point becomes fully
unstable, and the critical behavior is associated with a novel two-cycle
attractor of saddle-point character. However, for more complex substitution
rules, such as the Rudin-Shapiro sequence of four letters (which is
known to mimic Gaussian random fluctuations in a sense \cite{queffelec}) there
appears a surprisingly rich structure in the four-dimensional parameter
space. Besides the expected symmetric fixed point, there are non-diagonal
fixed points and continuous lines of two-cycle attractors, associated
with a marginal operator, which might give rise to non-universal critical
exponents. A scaling analysis indicates that this structure is responsible
for a novel critical universality class. In order to check these results,
we resorted to an independent thermodynamic calculation, on the basis
of a transfer matrix scheme.

The layout of this paper is as follows. In Section 2, we define the
polymer model, and present some renormalization-group calculations
for two- and four-letter (Rudin-Shapiro) substitution rules. We show
the existence of some surprising structures in parameter space, and, 
in particular, discuss a number of scaling results for the critical 
behavior. We then proceed to formulate an extension of the transfer 
matrix scheme for a two-polymer model. Although this technique has 
already been used for spin systems, its extension for this new situation 
requires a considerable amount of analytical work, as described in Sections
4 and 5. In Section 3, we just formulate the transfer-matrix scheme
for a two-polymer model. In Section 4, we study in great detail the
algebraic structure of the transfer matrices, in order to unveil, 
in Section 5, the existence of a recursion relation for the eigenvalues
of the transfer matrices associated with two successive generations of the 
hierarchical structure. In Section 6, we use the recursion relations in 
order to write down explicit thermodynamic functions. In Section 7 
we present the results for aperiodic models, which are compared to  the 
renormalization-group predictions.

\section{The interacting polymer model}

The binding-unbinding phase transition in a disordered model of two
directed and interacting polymers on a hierarchical lattice has been
investigated by Mukherji and Bhattacharjee \cite{mukherji}, and we follow 
these authors on the definition of the model. We simply place two directed 
polymers on a diamond hierarchical. They start at one end of the lattice and 
stretch continuously to the other end (the root sites). There is an attractive 
interaction, \( -\epsilon  \), whenever a bond of the lattice is
shared by a monomer of each polymer. This energy can be made to depend
on the position of the bond along a branch, in a random or deterministic fashion. 
Note that, although seemingly artificial, the model is nothing else than 
a \emph{bona-fide} Migdal-Kadanoff approximation for the same interacting problem 
on a genuine Bravais lattice.

In the basic cell of a diamond lattice, with \( q \) branches and
\( p \) bonds per branch, there are configurations of energy \( -p\epsilon \),
where the two polymers occupy the same \( p \) bonds of a branch,
and configurations of zero energy, where the polymers stretch along
different branches. Using the Boltzmann factor 
\( y=\exp \left( \beta \epsilon \right)  \)
and the combinatorial coefficient \( C_{q}^{2}=q\left( q-1\right) /2 \),
it is easy to write the RG recursion relation\begin{equation}
\label{eq1001}
y^{\prime }=\frac{1}{q}y^{p}+\frac{q-1}{q}.
\end{equation}
Taking \( p=2 \), for example, we see that besides the trivial fixed
points, \( y^{\ast }=1 \) and \( \infty  \), associated with zero
and infinite temperatures, there is a nontrivial fixed point, \( y^{\ast }=q-1 \),
which is physically acceptable for \( q>2 \) (there is no phase transition
on the simple diamond lattice with \( q=2 \) branches). Also, from
the linearization of the recursion relation about the nontrivial fixed
point, we have the thermal eigenvalue \begin{equation}
\label{eq1002}
\Lambda =\frac{2\left( q-1\right) }{q}.
\end{equation}

In order to obtain the specific-heat exponent, \( \alpha \), one
should note that, as the polymers are one-dimensional objects, the
thermodynamic extensivity of this model relates to the polymer length,
instead of the volume. Thus, the important density is the free energy
per monomer, which is assumed to behave according to the fundamental
scaling relation \begin{equation}
f\left( y^{\prime }\right) =pf\left( y\right) .
\end{equation}
From this relation, we have the critical exponent associated with
the specific heat, \begin{equation}
\label{alpha}
\alpha =2-\frac{\ln p}{\ln \Lambda }.
\end{equation}
This quantity will be used to characterize the possible universality
classes of the model, depending on \( q \), \( p \) and the presence
of aperiodically distributed interactions.

Consider, for example, for \( p=2 \), a layered distribution of interactions
\cite{haddad1}, \( \epsilon _{a} \) and \( \epsilon _{b} \), chosen
according to the two-letter period-doubling sequence,
\begin{equation}
\label{eq2.0}
a\rightarrow ab,\qquad b\rightarrow aa.
\end{equation}
In Figure 1 we give an example of the construction of a simple diamond lattice
with this kind of aperiodicity, starting from the letter \( a \).
Along each branch the interaction energies are distributed according
to the letters of the aperiodic sequence generated by the recursive application 
of the rule. The same arguments as used in the last paragraph to derive 
Eq. (\ref{eq1001}) for the uniform case, lead to a pair of recursion relations, 
\begin{equation}
\label{eq2.1}
y_{a}^{\prime }=\frac{1}{q}y_{a}y_{b}+\frac{q-1}{q},
\end{equation}
 and \begin{equation}
\label{eq2.2}
y_{b}^{\prime }=\frac{1}{q}y_{a}^{2}+\frac{q-1}{q},
\end{equation}
 where \( y_{a,b}=\exp \left( \beta \epsilon _{a,b}\right) \), and
\( \epsilon _{a,b}>0 \) is the interaction energy at bonds of types
\( a \) and \( b \), respectively. For \( y_{a}=y_{b}=y \), we
recover the recursion relation associated with the uniform model.

\begin{figure}
\begin{center}
\epsfbox{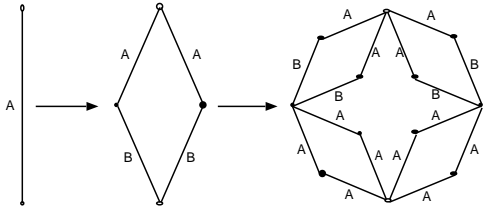}
\end{center}

\caption{Initial stages of the construction of a Migdal-Kadanoff hierarchical
lattice with \( q=2 \) branches and \( p=2 \) bonds per branch, and layered 
aperiodic interactions according to the period doubling rule,
\( a\rightarrow ab \), \( b\rightarrow aa \) (letters \( a \) and \( b \)
indicate the two possible values of the interaction energy, \( \epsilon_{a} \)
and \( \epsilon_{b} \)).}
\end{figure}

It is easy to see that, for \( q>2 \), there is no physically acceptable
nontrivial fixed point, except the symmetric fixed point, 
\( y_{a,b}^{\ast }=y^{\ast }=q-1 \).
The linearization of the recursion relations (\ref{eq2.1}) and (\ref{eq2.2})
in the neighborhood of this symmetric fixed point leads to the matrix
form \begin{equation}
\label{eq2.4}
\left( \begin{array}{c}
\Delta y_{a}^{\prime }\\
\Delta y_{b}^{\prime }
\end{array}\right) =\frac{y^{\ast }}{q}\left( \begin{array}{cc}
1 & 1\\
2 & 0
\end{array}\right) \left( \begin{array}{c}
\Delta y_{a}\\
\Delta y_{b}
\end{array}\right) ,
\end{equation}
with eigenvalues \( \Lambda _{1}=2y^{\ast }/q=2\left( q-1\right) /q \),
and \( \Lambda _{2}=-y^{\ast }/q=-(q-1)/q \). Therefore, if \( q>2 \),
we have \( \Lambda _{1}>1 \) and \( \left| \Lambda _{2}\right| <1 \),
which shows that the geometric fluctuations are completely irrelevant
in this case (since \( \Lambda _{1} \) is the same eigenvalue associated
with the non-trivial fixed point of the underlying uniform model,
and the behavior is irrelevant along the other direction).

We now turn to a more interesting case. Consider a diamond lattice
with \( p=3 \) bonds and \( q \) branches. Suppose that the (layered)
interactions are chosen according to a period-\( 3 \) two-letter
sequence, \( a\rightarrow abb \) and \( b\rightarrow aaa \). The
new recursion relations are given by \begin{equation}
\label{eq2.5}
y_{a}^{\prime }=\frac{1}{q}y_{a}y_{b}^{2}+\frac{q-1}{q},
\end{equation}
 and \begin{equation}
\label{eq2.6}
y_{b}^{\prime }=\frac{1}{q}y_{a}^{3}+\frac{q-1}{q}.
\end{equation}
 For \( q>3 \), there is just a single nontrivial fixed point, at
a symmetric location, \begin{equation}
\label{eq2.7}
y_{a}^{\ast }=y_{b}^{\ast }=y^{\ast }=-\frac{1}{2}+\frac{1}{2}\sqrt{4q-3}.
\end{equation}
 The linearization in the neighborhood of this fixed point leads to
the matrix equation\begin{equation}
\label{eq2.8}
\left( \begin{array}{c}
\Delta y_{a}^{\prime }\\
\Delta y_{b}^{\prime }
\end{array}\right) =\frac{\left( y^{\ast }\right) ^{2}}{q}\left( \begin{array}{cc}
1 & 2\\
3 & 0
\end{array}\right) \left( \begin{array}{c}
\Delta y_{a}\\
\Delta y_{b}
\end{array}\right) ,
\end{equation}
 with eigenvalues \begin{equation}
\label{eq2.9}
\Lambda _{1}=3\frac{y^{\ast 2}}{q}=\frac{3}{2q}\left[ 2q-1-\sqrt{4q-3}\right] ,
\end{equation}
 and \begin{equation}
\label{eq2.10}
\Lambda _{2}=-2\frac{y^{\ast 2}}{q}=-\frac{1}{q}\left[ 2q-1-\sqrt{4q-3}\right] .
\end{equation}
For \( 3<q<3+\sqrt{5} \), it is easy to see that \( \Lambda _{1}>1 \),
and \( \left| \Lambda _{2}\right| <1 \). As in the case of the simple
diamond lattice with \( p=2 \) bonds, geometric fluctuations are
irrelevant and the critical behavior is identical to the uniform case.
However, for \( q>3+\sqrt{5}=5.236068... \), we have 
\( \left| \Lambda _{2}\right| >1 \),
and the symmetric fixed point becomes fully unstable, and, therefore,
cannot be reached from arbitrary initial conditions (the values of
\( \epsilon _{a} \) and \( \epsilon _{b} \)). For example, for \( q=5 \),
we have \( y_{a}^{\ast }=y_{b}^{\ast }=y^{\ast }=1.561552... \),
with eigenvalues \( \Lambda _{1}=2.140568... \) and \( \Lambda _{2}=0.951363...<1 \).
For \( q=6 \), however, we have 
\( y_{a}^{\ast }=y_{b}^{\ast }=y^{\ast }=1.791287... \),
with eigenvalues \( \Lambda _{1}=2.573958... \) and \( \Lambda _{2}=1.143981...>1 \).
But, as in the case of spin models on hierarchical lattices \cite{haddad3},
there is a two-cycle in parameter space. It is easy to numerically
locate this cycle at 
\( \left( y_{a}^{\ast },y_{b}^{\ast }\right) =\left( 1.419001...,2.267305...\right) \)
and \( \left( 2.049103...,1.309541...\right)  \), with eigenvalues
of the linearized second iterate given by \( \Lambda _{1}=2.624300...>1 \)
and \( \Lambda _{2}=0.772598...<1 \). A new critical universality
class is therefore expected to be defined by this attractor.

The behavior in parameter space is much more interesting if we consider
the Rudin-Shapiro, four-letter substitution rule, \( a\rightarrow ac \),
\( b\rightarrow dc \), \( c\rightarrow ab \), \( d\rightarrow db \).
Consider a simple diamond lattice with \( p=2 \) bonds and \( q \)
branches. It is straightforward to write four algebraic recursion
relations,\begin{equation}
\label{eq2.11}
y_{a}^{\prime }=y_{a}y_{c}/q+\left( q-1\right) /q,\qquad y_{b}^{\prime }=
y_{d}y_{c}/q+\left( q-1\right) /q,
\end{equation}
\begin{equation}
\label{eq2.12}
y_{c}^{\prime }=y_{a}y_{b}/q+\left( q-1\right) /q,\qquad y_{d}^{\prime }=
y_{d}y_{b}/q+\left( q-1\right) /q,
\end{equation}
 which lead to the symmetric fixed point \begin{equation}
\label{eq2.13}
y_{a}^{\ast }=y_{b}^{\ast }=y_{c}^{\ast }=y_{d}^{\ast }=y^{\ast }=q-1.
\end{equation}
From the linearization about this fixed point, we have the 
eigenvalues\begin{equation}
\label{eq2.14}
\Lambda _{1}=2\frac{q-1}{q},\quad \Lambda _{2}=\sqrt{2}\frac{q-1}{q},\quad \Lambda _{3}
=-\sqrt{2}\frac{q-1}{q},\quad \Lambda _{4}=0.
\end{equation}
 The introduction of aperiodic interactions becomes relevant for the
simple diamond lattice if \( q>2+\sqrt{2}=3.41... \), which corresponds
to \( \left| \Lambda _{2}\right| =\left| \Lambda _{3}\right| >1 \).

Recursion relations (\ref{eq2.11}) and (\ref{eq2.12}) are so simple
that we can perform a number of detailed calculations. In particular,
it is easy to show the existence of additional, non-diagonal fixed
points, given by\begin{equation}
\label{eq2.15}
y_{a,d}^{\ast }=\frac{1}{4\left( q-1\right) }\left[ q\left( q^{2}-2q+2\right) 
\pm q\sqrt{\left( q^{2}-2\right) \left( q^{2}-4q+2\right) }\right] ,
\end{equation}
\begin{equation}
\label{eq2.16}
y_{b,c}^{\ast }=q-\frac{q-1}{y_{d,a}^{\ast }}.
\end{equation}

The Jacobian matrix associated with the linear form in the neighborhood
of these fixed points can be written as \begin{equation}
\label{eq2.17}
\frac{1}{q}\left( \begin{array}{cccc}
y_{c}^{\ast } & 0 & y_{a}^{\ast } & 0\\
0 & 0 & y_{d}^{\ast } & y_{c}^{\ast }\\
y_{b}^{\ast } & y_{a}^{\ast } & 0 & 0\\
0 & y_{d}^{\ast } & 0 & y_{b}^{\ast }
\end{array}\right) .
\end{equation}
Besides two trivial eigenvalues, \( \Lambda _{3}=0 \) and \( \Lambda _{4}=-1 \),
there is an additional pair of eigenvalues (\( \left| \Lambda _{1}\right| >1 \)
and \( \left| \Lambda _{2}\right| <1 \)) given by the solutions of
the quadratic equation\begin{equation}
\label{eq2.18}
2q^{4}\Lambda ^{2}-4q^{2}\left( q^{2}+q-1\right) \Lambda +q^{4}+8q^{3}-4q^{2}-8q+4=0.
\end{equation}
As \( \left| \Lambda _{4}\right| =1 \), we have a typical case of
marginal behavior, which cannot be analysed without resorting to higher-order
calculations. However, the marginal operator does not give rise to
a line of fixed points, as could be expected. Instead, with an additional
algebraic effort, it is possible to show the existence of a continuous
line whose points are two-cycles, by solving the polynomial 
equations\begin{equation}
\label{eq2.19}
y_{a,b,c,d}=y_{a,b,c,d}^{\prime \prime }\left( y_{a},y_{b},y_{c},y_{d}\right) ,
\end{equation}
where \( y^{\prime \prime } \) is the second iterate of the recursion
relations. Given any \( q>2+\sqrt{2} \), these equations lead to
a pair of one-parameter algebraic curves, which meet smoothly and
form a single, non-intersecting closed curve, containing the non-diagonal
fixed points. Any point belonging to this closed curve is mapped into
another point on the curve upon one iteration of the recursion relations,
and back to itself upon a further one. We have used algebraic computation
to check this result very thoroughly. As an example, for \( q=4 \),
we have the equations\begin{equation}
y_{a}=3\frac{t+4}{4t-3},
\end{equation}
\begin{equation}
y_{b}=-3\frac{70t^{2}+71t-114\pm g(t)}{t\left[ 6t^{2}-137t+78\pm g(t)\right] },
\end{equation}
\begin{equation}
y_{c}=-\frac{6t^{2}-128t+114\pm g(t)}{3t\left( t+4\right) }
\end{equation}
and \begin{equation}
y_{d}=t,
\end{equation}
with\begin{equation}
g(t)=(36t^{4}-2220t^{3}+15529t^{2}-27132t+12296)^{1/2},
\end{equation}
and the parameter \( t \) taking values between \( 1.5142\ldots  \)
and \( 5.4112\ldots  \). In Figure 2 we show a three-dimensional projection
of this attractor. The linearization of the second iterates about
any point of the curve leads to the eigenvalues \( 0 \), \( -1 \),
and a conjugate pair, \( \left| \Lambda _{1}\right| >1 \) and 
\( \left| \Lambda _{2}\right| <1 \)
(the values of which do not depend on the point about which linearization
is being carried out). The effects of the marginal eigenvalues on
the specific-heat exponent, related to the existence of an extended
attractor in parameter space, have to be checked very carefully, so
we turn to the direct thermodynamic analysis of the free-energy singularity.

\begin{figure}
\begin{center}
\epsfbox{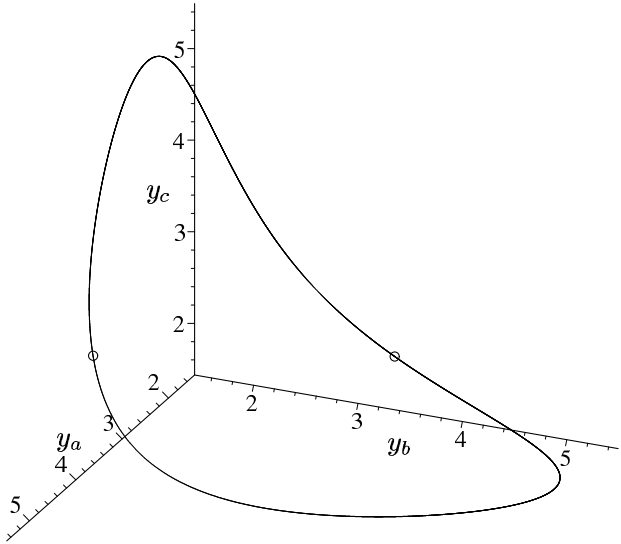}
\end{center}

\caption{Three-dimensional projection of the attractor of the RG recursion 
relations for the Rudin-Shapiro aperiodic sequence, in a lattice with 
$p=2$ and $q=4$. The non-diagonal fixed points are shown as circles.}
\end{figure}

Before proceeding to the transfer-matrix calculations, however, it is worth
remarking that, from a broad renormalization-group perspective, relevance and
irrelevance of aperiodic distributions of couplings is related to the
existence or not of a second eigenvalue with modulus larger than unity. The
general structure of the recursion relations can be used to derive a relevance
criterion which is ultimately based on the geometry of the lattice (that is,
the values of \( p \) and \( q \)), and some measure of the ``strength'' of the 
aperiodic fluctuations, such as the wandering exponent \cite{luck}. 

\section{Transfer matrix formulation}

One of us has successfully used a transfer matrix (TM) technique to
obtain the thermodynamic properties of several spin models on fractal
lattices \cite{andrade1,andrade2}. The essential step of this scheme
consists in the derivation of maps relating the eigenvalues of the
transfer matrices associated with two subsequent generations, \( G \)
and \( G+1 \). In a certain sense, this formalism is equivalent to
a method used by Derrida \emph{et al.} \cite{derrida} to establish
a map for the free energy, although it enables the calculation of
a correlation length, which turns out to be very useful to locate
the critical temperature (in spin systems).

In order to introduce the transfer matrix formulation, let us define
the model in a mathematically more precise way. Again, we will consider
a simple diamond hierarchical lattice (remembering that it has \( q \)
branches and \( p \) bonds per branch in each basic cell). At a generation
\( G \) of the hierarchical construction, the lattice has \( q^{G} \)
branches, each of them formed by \( p^{G} \) bonds. A polymer \( A, \)
extending from one root site to the other, is formed by \( p^{G} \)
connected monomers, and represented by a directed continuous path
between the two root sites. A given monomer, 
labeled \( i \) (\( 0\leq i\leq p^{G}-1 \)),
occupies one of the \( q^{G} \) available branches at the \( i \)th
position along the path. We may define a numbering for the \( q^{G} \)
branches of the lattice, and let the variables \( a_{i} \) indicate
which one is occupied by the \( i \)th monomer; clearly, 
\( 1\leq a_{i}\leq q^{G} \),
with analogous definitions for the other polymer, \( B \). The two
polymers interact at position \( i \), with energy \( -\epsilon _{i}<0 \),
if the \( i \)th monomers of the two distinct polymers occupy the
same bond (note that the energy depends only on the position \( i \),
since we will consider layered interactions). If the \( i \)th monomers occupy 
distinct bonds, the interaction energy vanishes.

This definition of the two-polymer model can be summarized by the
Hamiltonian

\begin{equation}
\label{eq1}
\mathcal{H}_{G}=-\sum _{i=0}^{p^{G}-1}\epsilon _{i}\delta \left( a_{i},b_{i}\right) ,
\end{equation}
where \( \delta \left( a_{i},b_{i}\right)  \) indicates a Kronecker
delta.

In Figure 3, we illustrate the simplest case, for \( G=1 \), \( p=2 \)
and \( q=2 \). Note that, in fact, we are considering a periodic
chain of \( N \) hierarchical cells, each one grown up to generation
\( G \). For a single cell, as each monomer can independently occupy
any of the \( q \) branches, there are \( q^{p} \) possible configurations
for each position \( i \) (remember that, as \( p=2 \), \( i \)
can have only two values in each cell), so that a total of \( 16 \)
possible states can be devised for this specific situation. However,
we have to take into account that each polymer is required to form
a continuous path between the root sites, so that several configurations
have to be excluded. Formally, we may calculate the partition function
including these forbidden configurations, if we introduce in the Hamiltonian
an additional term of the form 
\( \eta \sum _{i}V_{i}(a_{i},a_{i+1},b_{i},b_{i+1}) \),
with \( \eta \rightarrow \infty  \), such that \( V_{i}=0 \) for
the acceptable configurations, and \( V_{i}>0 \) whenever \( a_{i} \),
\( a_{i+1} \) and \( b_{i} \), \( b_{i+1} \) are not properly constrained.
The explicit form of this potential is somewhat cumbersome, and will
not be given here.

\begin{figure}
\begin{center}
{\centering \resizebox*{7cm}{1.8cm}{\includegraphics{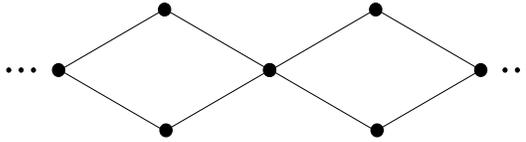}} \par}
\end{center}

\caption{Periodic chain of $N$ hierarchical cells joined by the root sites, each 
consisting on the first generation of a diamond lattice with $p=2$ and $q=2$. 
This is the initial stage of the construction of the transfer-matrix formalism 
for the interacting polymer model.}
\end{figure}

Although it is straightforward to write a partition function for the
particular case illustrated in Figure 3, \begin{equation}
\label{eq2}
Z_{1}=\left( 2e^{\beta \epsilon _{0}}e^{\beta \epsilon _{1}}+2\right) ^{N},
\end{equation}
the calculation of \( Z_{G} \), for arbitrary values of \( G \),
and hence of the thermodynamic properties of the model, represents
a much more difficult task. This is the reason to invoke the transfer
matrix technique. However, this problem of interacting polymers, with
interaction energies depending on monomer positions along the bonds,
requires a completely new definition of the transfer matrices.

For the sake of simplicity, we restrict the presentation of the formalism
to the homogeneous (uniform) system, which requires a smaller number
of different types of elementary TM's. It is straightforward to work
out an extension for the more complex situation of a model with aperiodic
interactions.

With the inclusion of the infinite-energy term, Eq. (\ref{eq1}) can
be written in the symmetrized form, \begin{equation}
\label{eq3}
\fl\mathcal{H}_{G}=-\frac{\epsilon }{2}\sum _{i=0}^{p^{G}-1}\left[ \delta 
\left( a_{i},b_{i}\right) +\delta \left( a_{i+1},b_{i+1}\right) \right] 
+\eta \sum _{i=0}^{p^{G}-1}V_{i}\left( a_{i},a_{i+1},b_{i},b_{i+1}\right),
\end{equation}
where we impose periodic boundary conditions, \( a_{p^{G}}=a_{0} \)
and \( b_{p^{G}}=b_{0} \). Now we define the \( q^{2G}\times q^{2G} \)
matrices, \begin{equation}
\fl(\mathbf{M}_{G}^{(i)})_{a_{i}b_{i};a_{i+1}b_{i+1}}=  
\exp \left\{ \frac{\beta \epsilon}{2}
\left[ \delta (a_{i},b_{i})+\delta (a_{i+1},b_{i+1})\right]
+\beta \eta V_{i}(a_{i},a_{i+1},b_{i},b_{i+1})\right\} ,\label{eq4} 
\end{equation}
and recall that the index \( i \) ranges from \( 0 \) to \( p^{G}-1 \)
(the length of each polymer in generation \( G \) is \( p^{G} \) ). 
Also, it should be remarked that, in the double indices of 
\( \mathbf{M}_{G}^{(i)} \), each term \( a_{i}b_{i} \) must be interpreted 
as a tensor product of the variables \( a_{i} \) and \( b_{i} \), 
each of them taking \( q^{G} \) independent values. The Boltzmann weights 
at each \( \mathbf{M}_{G}^{(i)} \) relate to the attracting energy between 
two monomers and to whether the \( i \)th monomer, placed at the bond 
\( a_{i}\left( b_{i}\right)  \), can be linked to the \( (i+1) \)th monomer 
at the bond \( a_{i+1}(b_{i+1}) \), without violating the continuity constraint. 
The definition of the TM's in Eq. (\ref{eq4}) leads to the formal 
identification of the partition function with the trace of a product 
of TM's, \begin{equation}
\label{eq5}
Z_{G}=\Tr \prod _{i=0}^{p^{G}-1}\mathbf{M}_{G}^{(i)} \equiv \Tr \mathbf{M}_{G}.
\end{equation}
Note that in this equation we considered \( N=1 \), since periodic boundary
conditions have already been enforced.

In the following  we will show that \( \mathbf{M}_{G} \) has just one
non-zero eigenvalue, which we call \( \eta_{G} \), and which is obviously
its trace. For the sake of clarity, we state now the result that \( \eta_{G} \)
is given by a recursion relation,\begin{equation}
\eta _{G}=q\left[ \eta _{G-1}^{2}+(q-1)\chi _{G-1}^{4}\right] ,
\end{equation}
with\begin{equation}
\chi _{G}=q^{\left( 2^{G}-1\right) },
\end{equation}
and \( \eta_{0}=y^{2} \). The next two sections are devoted to the derivation
of this result, and some readers may be interested on skipping directly to
Section 6, where we establish recursion relations for the relevant
thermodynamic functions.

\section{Detailed structure of the matrices}

The essential difficulty in the evaluation of \( Z_{G} \) has been
shifted into the calculation of the eigenvalues of a product of different 
kinds of TM's, \( \mathbf{M}_{G}^{(i)}. \) In order to emphasize 
the main ideas of the method, let us explore in greater detail the structure 
of the matrices in the case illustrated in Figure 3, where now the partition
function is just the one given by Eq. (\ref{eq2}), with 
\( \epsilon _{1}=\epsilon _{2} \) (remember we are considering uniform 
interactions for the time being). Defining now 
\( y=\exp \left( \beta \epsilon /2\right) , \)
we note that the TM's \( \mathbf{M}_{G=1}^{(i)} \) assume the two distinct
forms, \begin{equation}
\label{eq6}
\fl\mathbf{M}_{1}^{(1)}=\left( \begin{array}{cccc}
y^{2} & y & y^{2} & y\\
y & 1 & y & 1\\
y^{2} & y & y^{2} & y\\
y & 1 & y & 1
\end{array}\right) =\left( \begin{array}{cc}
1 & 1\\
1 & 1
\end{array}\right) \otimes \left( \begin{array}{cc}
y^{2} & y\\
y & 1
\end{array}\right) \equiv \mathbf{L}_{1}^{(1)}\otimes \mathbf{J}_{1}^{(1)},
\end{equation}
and

\begin{equation}
\label{eq7}
\fl\mathbf{M}_{1}^{(0)}=\left( \begin{array}{cccc}
y^{2} & 0 & 0 & 0\\
0 & 1 & 0 & 0\\
0 & 0 & y^{2} & 0\\
0 & 0 & 0 & 1
\end{array}\right) =\left( \begin{array}{cc}
1 & 0\\
0 & 1
\end{array}\right) \otimes \left( \begin{array}{cc}
y^{2} & 0\\
0 & 1
\end{array}\right) \equiv \mathbf{L}_{1}^{(0)}\otimes \mathbf{J}_{1}^{(0)},
\end{equation}
depending on whether the bonds at sites \( i \) and \( i+1 \) meet
at single vertex, where all 4 bonds are connected, or at two vertices,
where the bonds are connected pairwise. Comparing Eqs. (\ref{eq6})
and (\ref{eq7}), we note that several matrix elements in Eq. (\ref{eq6}),
which are equal to \( y \) and \( 1 \), are replaced by \( 0 \)'s
in Eq. (\ref{eq7}). They result from the presence of the term 
\( \exp (-\beta \eta )\rightarrow 0 \)
in the Boltzmann weights, indicating that the polymer that arrives
at a vertex where only two bonds meet cannot jump to a bond not incident
to that vertex. For this simple situation, Eq. (\ref{eq5}) reduces
to\begin{equation}
\label{eq8}
Z_{G=1}=\Tr  \left( \mathbf{M}_{1}^{(1)}\mathbf{M}_{1}^{(0)}\right) = 
\Tr \left( \mathbf{L}_{1}^{(1)}\mathbf{L}_{1}^{(0)}\otimes \mathbf{J}_{1}^{(1)}
\mathbf{J}_{1}^{(0)}\right) .
\end{equation}
The four eigenvalues of \( \mathbf{M}_{1}=\mathbf{M}_{1}^{(1)}\mathbf{M}_{1}^{(0)} \)
are given by the independent products of the eigenvalues of 
\( \mathbf{L}_{1}^{(1)}\mathbf{L}_{1}^{(0)} \)
(which are \( 2 \) and \( 0 \)) and those of \( \mathbf{J}_{1}^{(1)}\mathbf{J}_{1}^{(0)} \)
(which are \( y^{4}+1 \) and \( 0). \) Thus, \( \mathbf{M}_{1}^{(1)}\mathbf{M}_{1}^{(0)} \)
has one single non-vanishing eigenvalue, which is \( \Lambda =2\left( y^{4}+1\right)  \).

Restricting the analysis to \( p=2, \) we can show that a similar
result is valid for any \( q, \) as \( \mathbf{M}_{1}^{(1)} \) and
\( \mathbf{M}_{1}^{(0)} \) are expressed in terms of Kronecker products
of matrices with the same structure as those which are present in
Eqs. (\ref{eq6}) and (\ref{eq7}). All elements of the \( q\times q \)
matrix \( \mathbf{L}_{1}^{(1)} \) are equal to unity; 
\( (\mathbf{J}_{1}^{(1)})_{1,1}=y^{2}, \)
\( (\mathbf{J}_{1}^{(1)})_{1,l}=(\mathbf{J}_{1}^{(1)})_{l,1}=y \)
for \( l=2 \),\ldots{},\( q \), while all other elements are set
to unity; \( \mathbf{L}_{1}^{(0)} \) is the identity \( q\times q \)
matrix, while \( \mathbf{J}_{1}^{(0)} \) is a diagonal matrix with
\( y^{2} \) in the first entry, and \( 1 \) along the rest of the
diagonal. The only non-vanishing eigenvalue is \( \Lambda =q\left( y^{4}+q-1\right)  \),
which confirms the previous result for \( q=2. \)

\begin{figure}
\begin{center}
{\centering \resizebox*{7cm}{2cm}{\includegraphics{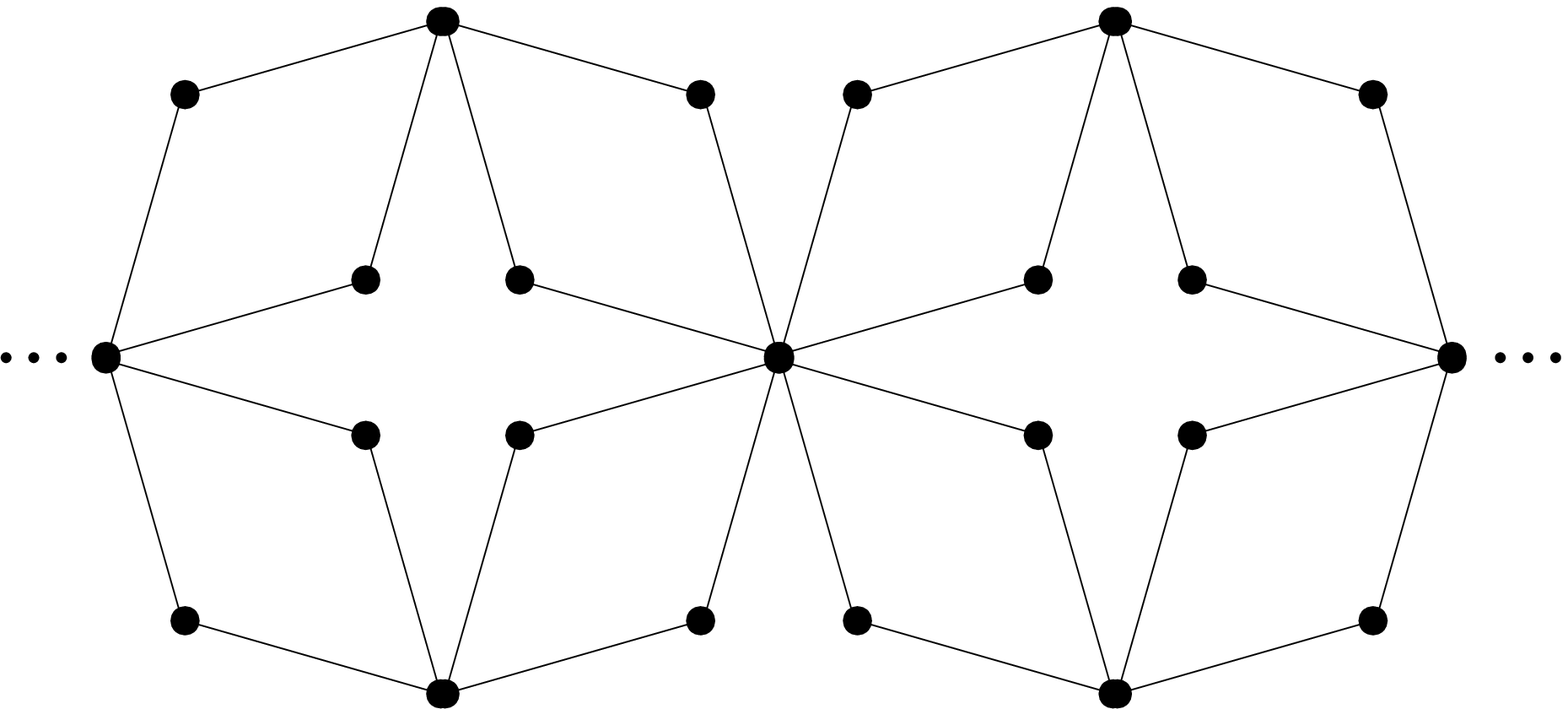}} \par}
\end{center}

\caption{Second stage of the construction of the transfer-matrix formalism. Now, 
each cell on the periodic chain is a diamond lattice of the second generation.}
\end{figure}

If we go into the next generation, \( G=2 \) (see Figure 4), the basic
cell of length \( 2^{2} \) is associated with three different types
of matrices, \( \mathbf{M}_{2}^{(j)} \), \( j=0,1,2 \). They describe,
respectively, the situations where the bonds at position \( i \)
meet with those at position \( i+1 \) at \( 1 \), \( 2 \) or \( 4 \)
distinct vertices, each one with connectivity \( 2q^{2}, \) \( 2q \)
and \( 2q^{0}. \) For \( q=2, \) these matrices are \begin{equation}
\label{eq9}
\mathbf{M}_{2}^{(2)}=\left( \begin{array}{cc}
1 & 1\\
1 & 1
\end{array}\right) \otimes \left( \begin{array}{cc}
1 & 1\\
1 & 1
\end{array}\right) \otimes \left( \begin{array}{cccc}
y^{2} & y & y & y\\
y & 1 & 1 & 1\\
y & 1 & 1 & 1\\
y & 1 & 1 & 1
\end{array}\right) ,
\end{equation}

\begin{equation}
\label{eq10}
\mathbf{M}_{2}^{(1)}=\left( \begin{array}{cc}
1 & 0\\
0 & 1
\end{array}\right) \otimes \left( \begin{array}{cc}
1 & 1\\
1 & 1
\end{array}\right) \otimes \left( \begin{array}{cccc}
y^{2} & y & 0 & 0\\
y & 1 & 0 & 0\\
0 & 0 & 1 & 1\\
0 & 0 & 1 & 1
\end{array}\right) ,
\end{equation}
 and

\begin{equation}
\label{eq11}
\mathbf{M}_{2}^{(0)}=\left( \begin{array}{cc}
1 & 0\\
0 & 1
\end{array}\right) \otimes \left( \begin{array}{cc}
1 & 0\\
0 & 1
\end{array}\right) \otimes \left( \begin{array}{cccc}
y^{2} & 0 & 0 & 0\\
0 & 1 & 0 & 0\\
0 & 0 & 1 & 0\\
0 & 0 & 0 & 1
\end{array}\right) .
\end{equation}
 The presence of \( 0 \)'s in Eqs. (\ref{eq10}) and (\ref{eq11}),
and the form of the matrices for larger values of \( q \) follow
from the same kind of arguments already used to discuss the generation
\( G=1. \) To evaluate the partition function \( Z_{2} \) it is
necessary to identify the order in which the factors 
\( \mathbf{M}_{2}^{(1)},\mathbf{M}_{2}^{(2)} \)
and \( \mathbf{M}_{2}^{(3)} \) are multiplied to form the matrix
\( \mathbf{M}_{2}. \) This can be easily realized if we recall the
association of the different matrix types with the various kinds of
vertices along the hierarchical lattice. We thus have

\begin{equation}
\label{eq12}
\mathbf{M}_{2}=\mathbf{M}_{2}^{(0)}\mathbf{M}_{2}^{(1)}\mathbf{M}_{2}^{(0)}
\mathbf{M}_{2}^{(2)},
\end{equation}
from which it is straightforward to calculate \( Z_{2} \).

Let us now obtain the structure of the matrices \( \mathbf{M}_{G} \)
for any \( G. \) This follows from Eq. (\ref{eq5}), from the hierarchical
structure of the lattice, and from the detailed discussion of the
form of matrices \( \mathbf{M}_{1}^{(i)} \) and \( \mathbf{M}_{2}^{(i)} \).
The first important property related to the structure of the lattice
is that, along each branch, there are sites with different connectivities,
and they appear according to a well-defined sequence. The difference
in connectivities results in local variations of degrees of freedom,
since the polymers can choose among different numbers of branches.
The inner sites can be of \( G \) types, the connectivities of which
are \( 2 \), \( 2q \), \( 2q^{2} \), \ldots{}, \( 2q^{G-1} \),
while the root sites have connectivity \( 2q^{G} \). We identify
the type of a site by the variable \( s \), such that the connectivity
of a particular site is \( 2q^{s} \). Let \( S_{G} \) be the sequence
of \( p^{G} \) numbers that identifies the order in which the several
kinds of sites appear along a branch in generation \( G. \) For instance,
\( S_{1}=\{0,1\} \) and \( S_{2}=\{0,1,0,2\} \). We note that \( S_{G} \)
can be obtained by the concatenation of two sequences \( S_{G-1} \),
replacing the last symbol \( (G-1) \) by \( G \). Also, we observe
that \( S_{G} \) contains \( 2^{G-1} \) symbols \( 0, \) \( 2^{G-2} \)
symbols \( 1 \), and so forth, until one single symbol \( G-1 \)
(at the central position) and one symbol \( G \) at the rightmost
position. As Eq. (\ref{eq12}) suggests, the matrix associated with
a site of type \( s \) is \( \mathbf{M}_{G}^{(s)}, \) so that \( \mathbf{M}_{G} \)
decomposes into a product of elementary matrices in a well defined
order,\begin{equation}
\label{eq13}
\mathbf{M}_{G}=\mathbf{M}_{G}^{(0)}\mathbf{M}_{G}^{(1)}\mathbf{M}_{G}^{(0)}
\mathbf{M}_{G}^{(2)}\mathbf{M}_{G}^{(0)}\mathbf{M}_{G}^{(1)}\mathbf{M}_{G}^{(0)}
\mathbf{M}_{G}^{(3)}...\mathbf{M}_{G}^{(0)}\mathbf{M}_{G}^{(G)},
\end{equation}

We now study the structure of the \( q^{2G}\times q^{2G} \) matrices
\( \mathbf{M}_{G}^{(i)}. \) Each of them can be expressed by the
Kronecker product of two \( q^{G}\times q^{G} \) matrices, \( \mathbf{L}_{G} \)
and \( \mathbf{J}_{G}, \) so that\begin{eqnarray}
\label{eq14}
\mathbf{M}_{G} & = & \left[ \mathbf{L}_{G}^{(0)}\mathbf{L}_{G}^{(1)}\mathbf{L}_{G}^{(0)}
\mathbf{L}_{G}^{(2)}\mathbf{L}_{G}^{(0)}\mathbf{L}_{G}^{(1)}\mathbf{L}_{G}^{(0)}
\mathbf{L}_{G}^{(3)}...\mathbf{L}_{G}^{(0)}\mathbf{L}_{G}^{(G)} \right] \nonumber \\
& & \otimes \left[ \mathbf{J}_{G}^{(0)}\mathbf{J}_{G}^{(1)}\mathbf{J}_{G}^{(0)}
\mathbf{J}_{G}^{(2)}\mathbf{J}_{G}^{(0)}\mathbf{J}_{G}^{(1)}\mathbf{J}_{G}^{(0)}
\mathbf{J}_{G}^{(3)}...\mathbf{J}_{G}^{(0)}\mathbf{J}_{G}^{(G)} \right] \nonumber \\
& \equiv & \mathbf{Q}_{G}\otimes \mathbf{R}_{G},
\end{eqnarray}
where the sequence \( S_{G} \) controls the numbering order in both
\( \mathbf{Q}_{G} \) and \( \mathbf{R}_{G} \). Then, we notice from
Eqs. (\ref{eq9}-\ref{eq11}) that each \( \mathbf{L}_{G}^{(i)} \)
can be further expressed by the Kronecker product of \( G \) matrices
of order \( q \), each of which is either the unit \( q\times q \)
matrix \( \mathbf{I} \) or the constant \( q\times q \) matrix \( \mathbf{K}, \)
with all elements set to unity.

The factor \( \mathbf{K} \) expresses allowed transitions of the
polymer from a given branch to neighboring ones, while \( \mathbf{I} \)
indicates a restriction for the polymer to change from a branch to
others. So, it is easy to see that \( \mathbf{L}_{G}^{(G)} \), which
describes the site with connectivity \( 2q^{G}, \) is formed by the
product of \( G \) matrices \( \mathbf{K}, \) while \( \mathbf{L}_{G}^{(0)} \),
related to sites with connectivity \( 2q^{0}, \) is formed by products
of matrices \( \mathbf{I} \) only. If we define \( \mathbf{A}^{\otimes G} \)
to be the Kronecker product of \( G \) matrices \( \mathbf{A}, \)
then we can write the general form of the matrices \( \mathbf{L}_{G}^{(g)} \)
as\begin{equation}
\label{eq15}
\mathbf{L}_{G}^{(g)}=\mathbf{I}^{\otimes (G-g)}\otimes \mathbf{K}^{\otimes g}.
\end{equation}
Also, it should be noted that, for \( g<G \), the matrices \( \mathbf{L}_{G}^{(g)} \)
and \( \mathbf{L}_{G-1}^{(g)} \) are related by\begin{equation}
\label{eq16}
\mathbf{L}_{G}^{(g)}=\mathbf{I}\otimes \mathbf{L}_{G-1}^{(g)}.
\end{equation}

The matrices \( \mathbf{J}_{G}^{(i)} \) cannot be further decomposed
in terms of Kronecker products, but they can be expressed as \begin{equation}
\label{eq17}
\mathbf{J}_{G}^{(g)}=\mathbf{L}_{G}^{(g)}+\mathbf{H}_{G}^{(g)},
\end{equation}
where \( \mathbf{H}_{G}^{(g)} \) is given by 
\begin{equation}
\label{eq18}
\left( \mathbf{H}^{(g)}_{G}\right) _{ij}=\cases {y^{2}-1& if $i=j=1$\\
y-1& if $i=1$,\rm{ } $j\in \left\{ 2,\ldots ,q^{g}\right\}$ \\
y-1& if $i\in \left\{ 2,\ldots ,q^{g}\right\}$ ,\rm{ } $j=1$\\
0 & otherwise.}.
\end{equation}
If \( g<G \), it is also possible to see that
\begin{equation}
\label{eq19}
\mathbf{H}_{G}^{(g)}=\mathbf{H}_{G-1}^{(g)}\oplus 0\oplus 0\oplus \ldots \oplus 0,
\end{equation}
where \( 0 \) indicates the null \( q^{G-1}\times q^{G-1} \) matrix,
which appears \( q-1 \) times in the direct sum. This expression
can also be written as \begin{equation}
\label{eq20}
\mathbf{H}_{G}^{(g)}=\left( \begin{array}{ccccc}
1 & 0 & 0 & \ldots  & 0\\
0 & 0 & 0 &  & 0\\
0 & 0 & 0 &  & 0\\
\vdots  &  &  & \ddots  & \vdots \\
0 & 0 & 0 & \ldots  & 0
\end{array}\right) \otimes \mathbf{H}_{G-1}^{(g)},
\end{equation}
where the first matrix is of order \( q \). Combining this last equation
with Eq. (\ref{eq16}), we obtain \begin{equation}
\label{eq21}
\fl\mathbf{J}_{G}^{(g)}=\mathbf{I}\otimes \mathbf{L}_{G-1}^{(g)}+\mathbf{H}_{G}^{(g)}=
\left( \begin{array}{ccccc}
\mathbf{J}_{G-1}^{(g)} & 0 & 0 & \ldots  & 0\\
0 & \mathbf{L}_{G-1}^{(g)} & 0 & \ldots  & 0\\
0 & 0 & \mathbf{L}_{G-1}^{(g)} & \ldots  & 0\\
\vdots  & \vdots  & \vdots  & \ddots  & \vdots \\
0 & 0 & 0 & \ldots  & \mathbf{L}_{G-1}^{(g)}
\end{array}\right) ,
\end{equation}
which is valid for \( g<G. \) In this expression, the \( 0 \)'s
represent null matrices of order \( q^{G-1} \), so that \( \mathbf{J}_{G}^{(g)} \)
is a block-diagonal matrix, with one \( \mathbf{J}_{G-1}^{(g)} \)
block and \( q-1 \) blocks \( \mathbf{L}_{G-1}^{(g)} \) in the diagonal.

The particular structure of the matrices \( \mathbf{L}_{G}^{(g)} \)
and \( \mathbf{J}_{G}^{(g)} \) leads to a recurrence relation for
the only non-zero eigenvalue of \( \mathbf{M}_{G} \) in terms of
the corresponding eigenvalue of \( \mathbf{M}_{G-1} \). This will
be shown in the next Section.

\section{Eigenvalues of the matrix \protect\( \mathbf{M}_{G}\protect \)}

The eigenvalues of \( \mathbf{M}_{G} \) are given by all distinct
products of the eigenvalues of \( \mathbf{Q}_{G} \) and \( \mathbf{R}_{G} \).
Let us first consider the eigenvalues of \( \mathbf{Q}_{G} \). According
to Eq. (\ref{eq14}), \( \mathbf{Q}_{G} \) is expressed by usual
matrix products of matrices which are themselves Kronecker products
of only two types of matrices, \( \mathbf{I} \) and \( \mathbf{K} \).
Then it is easy to show that\begin{equation}
\label{eq22}
\mathbf{Q}_{G}=\mathbf{K}\otimes \mathbf{K}^{2}\otimes \mathbf{K}^{4}\otimes \ldots 
\otimes \mathbf{K}^{2^{G-2}}\otimes \mathbf{K}^{2^{G-1}}.
\end{equation}
Using the relation \begin{equation}
\label{eq23}
\mathbf{K}^{n}=q^{n-1}\mathbf{K},
\end{equation}
and the identity\begin{equation}
\label{eq25}
\sum _{g=1}^{G-1}\left( 2^{g}-1\right) =2^{G}-G-1,
\end{equation}
it is possible to write Eq. (\ref{eq22}) as \begin{equation}
\label{eq24}
\mathbf{Q}_{G}=\left( \prod _{g=0}^{G-1}q^{2^{g}-1}\right) \mathbf{K}^{\otimes G}.
\end{equation}
The rank of \( \mathbf{K} \) is unity, and its only one non-zero
eigenvalue is \( q. \) It follows that \( q^{G} \) is the only non-zero
eigenvalue of \( \mathbf{K}^{\otimes G} \), so that \( \chi _{G} \),
the eigenvalue of \( \mathbf{Q}_{G} \), is given by \begin{equation}
\label{eq26}
\chi _{G}=q^{\left( 2^{G}-1\right) }.
\end{equation}

Now, let us calculate the eigenvalues of \( \mathbf{R}_{G} \), defined
through Eq. (\ref{eq14}). First, we write \( \mathbf{R}_{G} \) in
the form \begin{equation}
\label{eq27}
\mathbf{R}_{G}=\prod _{l=1}^{2^{G}}\mathbf{J}_{G}^{\left( S_{G}\right) _{l}}
=\left( \prod _{l=1}^{2^{G}-1}\mathbf{J}_{G}^{\left( S_{G}\right) _{l}}\right) 
\mathbf{J}_{G}^{\left( G\right) },
\end{equation}
where \( \left( S_{G}\right) _{l} \) represents the \( l \)th number
in the sequence \( S_{G} \). Then, we note that the matrix 
\( \mathbf{J}_{G}^{\left( G\right) } \)
can be written as \begin{equation}
\label{eq28}
\mathbf{J}_{G}^{\left( G\right) }=\left( \begin{array}{ccccc}
\mathbf{J}_{G-1}^{\left( G-1\right) } & \mathbf{E}_{G-1} & \mathbf{E}_{G-1} & \ldots  & 
\mathbf{E}_{G-1}\\
\mathbf{F}_{G-1} & \mathbf{L}_{G-1}^{\left( G-1\right) } & \mathbf{L}_{G-1}^{\left( G-1\right) } &
 \ldots  & \mathbf{L}_{G-1}^{\left( G-1\right) }\\
\mathbf{F}_{G-1} & \mathbf{L}_{G-1}^{\left( G-1\right) } & \mathbf{L}_{G-1}^{\left( G-1\right) } &
 \ldots  & \mathbf{L}_{G-1}^{\left( G-1\right) }\\
\vdots  & \vdots  & \vdots  & \ddots  & \vdots \\
\mathbf{F}_{G-1} & \mathbf{L}_{G-1}^{\left( G-1\right) } & \mathbf{L}_{G-1}^{\left( G-1\right) } &
 \ldots  & \mathbf{L}_{G-1}^{\left( G-1\right) }
\end{array}\right) ,
\end{equation}
where all elements of the \( q^{G-1}\times q^{G-1} \) matrix \( \mathbf{E}_{G-1} \)
are equal to unity, with exception of those of the first row, which
are equal to \( y \). The matrix \( \mathbf{F}_{G-1} \) is the transpose
of \( \mathbf{E}_{G-1}. \) For \( g<G \), we recall that Eq. (\ref{eq21})
uncovers the block-diagonal structure of \( \mathbf{J}_{G}^{(g)} \),
so that \begin{equation}
\label{eq29}
\fl\prod _{l=1}^{2^{G}-1}\mathbf{J}_{G}^{\left( S_{G}\right) _{l}}=
\left( \begin{array}{cccc}
\prod _{l=1}^{2^{G}-1}\mathbf{J}_{G}^{\left( S_{G}\right) _{l}} & 0 & \ldots  & 0\\
0 & \prod _{l=1}^{2^{G}-1}\mathbf{L}_{G}^{\left( S_{G}\right) _{l}} & \ldots  & 0\\
\vdots  & \vdots  & \ddots  & \vdots \\
0 & 0 & \ldots  & \prod _{l=1}^{2^{G}-1}\mathbf{L}_{G}^{\left( S_{G}\right) _{l}}
\end{array}\right) ,
\end{equation}
with \( q \) blocks of order \( q^{G-1} \) in the diagonal. Let
us now introduce the notation \begin{equation}
\label{eq30}
\Pi _{\mathbf{J}}=\prod _{l=1}^{2^{G}-1}\mathbf{J}_{G-1}^{\left( S_{G}\right) _{l}},
\end{equation}
with the analogous definition for the product of the matrices 
\( \mathbf{L}_{G}^{\left( g\right) } \).
Making use of Eqs. (\ref{eq28}) and (\ref{eq29}), we obtain \begin{equation}
\fl\mathbf{R}_{G}=\left( \begin{array}{ccccc}
\Pi _{\mathbf{J}}\mathbf{J}_{G-1}^{\left( G-1\right) } & \Pi _{\mathbf{J}}\mathbf{E}_{G-1} & 
\Pi _{\mathbf{J}}\mathbf{E}_{G-1} & \ldots  & \Pi _{\mathbf{J}}\mathbf{E}_{G-1}\\
\Pi _{\mathbf{L}}\mathbf{F}_{G-1} & \Pi _{\mathbf{L}}\mathbf{L}_{G-1}^{\left( G-1\right) } & 
\Pi _{\mathbf{L}}\mathbf{L}_{G-1}^{\left( G-1\right) } & \ldots  & 
\Pi _{\mathbf{L}}\mathbf{L}_{G-1}^{\left( G-1\right) }\\
\Pi _{\mathbf{L}}\mathbf{F}_{G-1} & \Pi _{\mathbf{L}}\mathbf{L}_{G-1}^{\left( G-1\right) } & 
\Pi _{\mathbf{L}}\mathbf{L}_{G-1}^{\left( G-1\right) } & \ldots  & 
\Pi _{\mathbf{L}}\mathbf{L}_{G-1}^{\left( G-1\right) }\\
\vdots  & \vdots  & \vdots  & \ddots  & \vdots \\
\Pi _{\mathbf{L}}\mathbf{F}_{G-1} & \Pi _{\mathbf{L}}\mathbf{L}_{G-1}^{\left( G-1\right) } & 
\Pi _{\mathbf{L}}\mathbf{L}_{G-1}^{\left( G-1\right) } & \ldots  & 
\Pi _{\mathbf{L}}\mathbf{L}_{G-1}^{\left( G-1\right) }
\end{array}\right) .
\end{equation}

Let us consider the entries of this matrix. Take, for instance,\begin{eqnarray}
\Pi _{\mathbf{L}}\mathbf{L}_{G-1}^{\left( G-1\right) } & = & 
\left( \prod _{l=1}^{2^{G}-1}\mathbf{L}_{G-1}^{\left( S_{G}\right) _{l}}\right) 
\mathbf{L}_{G-1}^{\left( G-1\right) }\nonumber \\
& = & \left( \prod _{l=1}^{2^{G-1}}\mathbf{L}_{G-1}^{\left( S_{G}\right) _{l}}\right) 
\left( \prod _{l=2^{G-1}+1}^{2^{G}-1}\mathbf{L}_{G-1}^{\left( S_{G}\right) _{l}}\right) 
\mathbf{L}_{G-1}^{\left( G-1\right) }.
\end{eqnarray}
The first factor is clearly \( \mathbf{Q}_{G-1} \), as the sequence
\( S_{G} \) is equivalent to \( S_{G-1} \) until the position \( l=2^{G-1} \).
However, \( S_{G} \) is also identical to \( S_{G-1} \) between
the positions \( l=2^{G-1}+1 \) and \( l=2^{G}-1 \), according to
the rule to generate \( S_{G} \) from \( S_{G-1} \). Thus, this
factor, multiplied by \( \mathbf{L}_{G-1}^{(G-1)} \), also leads
to \( \mathbf{Q}_{G-1} \), \begin{equation}
\label{eq33}
\Pi _{\mathbf{L}}\mathbf{L}_{G-1}^{\left( G-1\right) }=\mathbf{Q}_{G-1}^{2}.
\end{equation}
Following the same arguments, it is possible to show that \begin{equation}
\label{eq34}
\Pi _{\mathbf{J}}\mathbf{J}_{G-1}^{\left( G-1\right) }=\mathbf{R}_{G-1}^{2},
\end{equation}
so that \begin{equation}
\label{eq35}
\mathbf{R}_{G}=\left( \begin{array}{ccccc}
\mathbf{R}_{G-1}^{2} & \Pi _{\mathbf{J}}\mathbf{E}_{G-1} & 
\Pi _{\mathbf{J}}\mathbf{E}_{G-1} & \ldots  & \Pi _{\mathbf{J}}\mathbf{E}_{G-1}\\
\Pi _{\mathbf{L}}\mathbf{F}_{G-1} & \mathbf{Q}_{G-1}^{2} & \mathbf{Q}_{G-1}^{2} & 
\ldots  & \mathbf{Q}_{G-1}^{2}\\
\Pi _{\mathbf{L}}\mathbf{F}_{G-1} & \mathbf{Q}_{G-1}^{2} & \mathbf{Q}_{G-1}^{2} & 
\ldots  & \mathbf{Q}_{G-1}^{2}\\
\vdots  & \vdots  & \vdots  & \ddots  & \vdots \\
\Pi _{\mathbf{L}}\mathbf{F}_{G-1} & \mathbf{Q}_{G-1}^{2} & \mathbf{Q}_{G-1}^{2} & 
\ldots  & \mathbf{Q}_{G-1}^{2}
\end{array}\right) .
\end{equation}

Now we recall that \( \mathbf{R}_{G} \) is a product of the matrices
\( \mathbf{J}_{G}^{(g)} \), including \( \mathbf{J}_{G}^{(G)} \).
The rank of this last matrix is unity, as one sees from Eq. (\ref{eq28}).
Using the Frobenius inequality for the rank of matrices \cite{matrix},
we then see that the rank of \( \mathbf{R}_{G} \) is also unity.
Then, the only non-zero eigenvalue \( \lambda _{G} \) equals the
trace of \( \mathbf{R}_{G}, \) so that \begin{equation}
\label{eq36}
\lambda _{G}=\Tr \mathbf{R}_{G-1}^{2}+(q-1)\Tr \mathbf{Q}_{G-1}^{2}.
\end{equation}
However, it is clear that \( \Tr\mathbf{R}_{G-1}^{2}\equiv \lambda _{G-1}^{2} \).
Also, from Eq. (\ref{eq26}), we have \( \Tr\mathbf{Q}_{G-1}^{2}\equiv  \)
\( \chi _{G-1}^{2} \), so that \begin{equation}
\label{eq37}
\lambda _{G}=\lambda _{G-1}^{2}+(q-1)\chi _{G-1}^{2}.
\end{equation}
As \( \lambda _{0}=y^{2} \), this equation gives rise to a recursion
relation for the eigenvalues of \( \mathbf{R}_{G} \). If we call
\( \eta _{G} \) the only non-zero eigenvalue of \( \mathbf{M}_{G} \),
we may write \begin{equation}
\label{eq38}
\eta _{G}=\chi _{G}\lambda _{G}=\chi _{G}\left[ \lambda _{G-1}^{2}+(q-1)\chi _{G-1}^{2}\right] .
\end{equation}
From Eq. (\ref{eq5}), one sees that \( Z _{G}=\eta _{G} \), since the trace is just
this non-zero eigenvalue. Now, using Eq. (\ref{eq26}), 
we have \begin{equation}
\label{eq39}
\chi _{G}=q\chi _{G-1}^{2},
\end{equation}
from which we are finally led to the map \begin{equation}
\label{eq40}
\eta _{G}=q\left[ \eta _{G-1}^{2}+(q-1)\chi _{G-1}^{4}\right] .
\end{equation}

\section{Thermodynamic functions}

If we take the Boltzmann constant \( k_{B}=1 \), the free energy
per monomer of the system may be written as \begin{equation}
f_{G}=-\frac{T}{2^{G}}\ln Z _{G}=-\frac{T}{2^{G}}\ln \eta _{G}.
\end{equation}
Defining an auxiliary map, \( K_{G}=\chi _{G}^{2}/\eta _{G} \), we have
\begin{equation}
\label{eq42}
f_{G}=f_{G-1}-\frac{T}{2^{G}}\ln \left( 1+(q-1)K_{G-1}^{2}\right) -\frac{T}{2^{G}}\ln 2,
\end{equation}
 where

\begin{equation}
\label{eq43}
K_{G}=\frac{qK_{G-1}^{2}}{1+(q-1)K_{G-1}^{2}}.
\end{equation}

The recursive iteration of Eqs. (\ref{eq42}) and (\ref{eq43}), with
the initial conditions \( f_{0}=-\epsilon  \) and \( K_{0}=\exp (-\beta \epsilon ) \),
leads to the free energy per monomer for any generation \( G \),
which converges to a well-defined free-energy in the thermodynamic
limit, \( G\rightarrow \infty  \). Maps for additional thermodynamic
functions, as the entropy and the specific heat, can be obtained by
taking the derivative of Eqs. (\ref{eq42}) and (\ref{eq43}) with
respect to temperature. For the entropy per monomer, for example,
we obtain \begin{equation}
\label{eq44}
\fl s_{G}=s_{G-1}+\frac{1}{2^{G}}\ln \left( 1+(q-1)K_{G-1}^{2}\right) 
+\frac{T}{2^{G}}\frac{2(q-1)K_{G-1}}{1+(q-1)K_{G-1}^{2}}\frac{\partial K_{G-1}}{\partial T}+
\frac{1}{2^{G}}\ln 2,
\end{equation}
where \begin{equation}
\label{eq45}
\frac{\partial K_{G}}{\partial T}=\frac{2qK_{G-1}}{\left( 1+(q-1)K_{G-1}^{2}\right) ^{2}}
\frac{\partial K_{G-1}}{\partial T}.
\end{equation}

\begin{figure}
\begin{center}
{\centering \resizebox*{7cm}{7cm}{\includegraphics{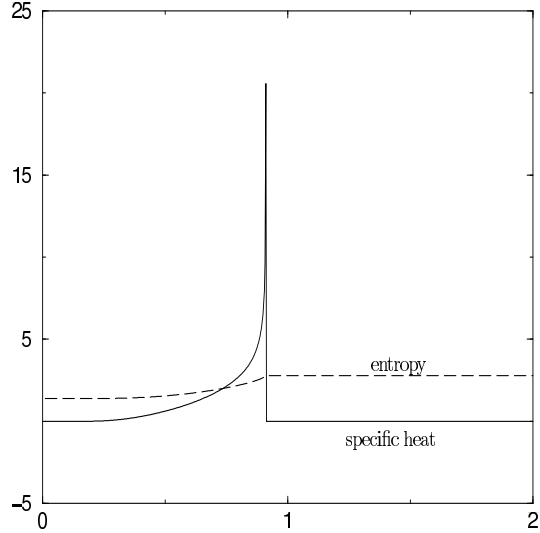}} \par}
\end{center}

\caption{Entropy and specific heat for a uniform model, in a lattice with $p=2$ and $q=4$,
calculated using the transfer-matrix technique. In this figure $\epsilon /k_{B}=1$.}
\end{figure}

As an example, in Figure 5 we show the results for the entropy
and specific heat of a uniform model on the lattice with \( p=2 \)
and \( q=4 \). Numerical analysis shows there is a genuine singularity, 
associated with a divergence of the specific heat at a critical temperature. 
Note the interesting behavior of the system above the transition temperature, 
with a constant entropy per monomer, and,
consequently, zero specific heat. On physical grounds, this result
should have been anticipated, since the polymers are completely unbound
on the high-temperature phase, and the maximum amount of disorder
is attained independently of temperature. The RG approach, however,
does not yield such a global picture of the thermodynamics, which
is possible only in the TM framework.

To check the reliability of the method, and its compatibility with
the RG results, we may compare the critical temperature and exponent
it yields with those predicted by the renormalization-group calculation.
For any value of \( q \), the fixed point \( y^{\ast }=q-1 \), with
\( y=\left( \exp \beta \epsilon \right)  \), gives the critical 
temperature\begin{equation}
T_{C}=\frac{\epsilon }{k_{B}}\frac{1}{\ln \left( q-1\right) },
\end{equation}
with a critical exponent given by Eqs. (\ref{alpha}) and (\ref{eq1002}).
For \( q=4 \) and \(\epsilon /k_{B}=1 \), we have \( T_{C}=0.910239\ldots  \) and 
\( \alpha =0.290488\ldots  \).The numerical analysis of data in Figure 4 leads to 
\( T_{C}\simeq 0.910239 \) and \( \alpha \simeq 0.2905 \), which confirms 
the accuracy of the method.

Now we remark that the models we are interested in include more than
one interaction energy, depending on the position \( i \) along the
path between the root sites. The basic steps of the TM scheme can
be adapted in order to obtain the corresponding maps, although much
attention has to be paid to all of the details. For instance, if we
consider an aperiodic model with the presence of two distinct interaction
energies, which are placed along the lattice according to the period-doubling
sequence \( a\rightarrow ab \), \( b\rightarrow aa \), the method
requires the definition of two matrices \( \mathbf{M}_{G}^{a} \)
and \( \mathbf{M}_{G}^{b} \), the eigenvalues of which are \( \eta _{G}^{(a)} \)
and \( \eta _{G}^{(b)} \). The maps for \( \eta _{G}^{(a)} \) and
\( \eta _{G}^{(b)} \) are written as \begin{equation}
\label{eq46}
\eta _{G}^{(a)}=q\left[ \eta _{G-1}^{(a)}\eta _{G-1}^{(b)}+(q-1)\chi _{G-1}^{4}\right] ,
\end{equation}
 and\begin{equation}
\label{eq47}
\eta _{G}^{(b)}=q\left[ \left( \eta _{G-1}^{(a)}\right) ^{2}+(q-1)\chi _{G-1}^{4}\right] .
\end{equation}
Note that each one of these eigenvalues gives rise to a different partition function, 
associated with the choice of \( a \) or \( b \) as the initial letter to be iterated
according to the inflation rule. We are always interested on sequences generated by the
recursive application of the rule to the initial letter \( a \).

If the aperiodicity is induced by the four-letter Rudin-Shapiro sequence,
\( a\rightarrow ac \), \( b\rightarrow dc \), \( c\rightarrow ab \),
\( d\rightarrow db \), the set of four maps for the eigenvalues are
given by \begin{equation}
\label{eq48}
\eta _{G}^{(a)}=q\left[ \eta _{G-1}^{(a)}\eta _{G-1}^{(c)}+(q-1)\chi _{G-1}^{4}\right] ,
\end{equation}
\begin{equation}
\label{eq49}
\eta _{G}^{(b)}=q\left[ \eta _{G-1}^{(c)}\eta _{G-1}^{(d)}+(q-1)\chi _{G-1}^{4}\right] ,
\end{equation}
\begin{equation}
\label{eq50}
\eta _{G}^{(c)}=q\left[ \eta _{G-1}^{(a)}\eta _{G-1}^{(b)}+(q-1)\chi _{G-1}^{4}\right] ,
\end{equation}
and\begin{equation}
\label{eq51}
\eta _{G}^{(d)}=q\left[ \eta _{G-1}^{(b)}\eta _{G-1}^{(d)}+(q-1)\chi _{G-1}^{4}\right] .
\end{equation}
The free energy per monomer, along with its temperature derivatives,
can be similarly defined, so that the singularity at the phase transition
can be analysed directly.

\section{Discussions and results}

As discussed in Sec. 2, the RG analysis of the uniform model indicates
the presence of a second-order phase transition for \( q>2 \), with
the specific-heat critical exponent given by Eq. (\ref{alpha}). In
case of irrelevant aperiodicity, that is, when the diagonal fixed
point has just one relevant eigenvalue, \( \alpha  \) is given by
the same expression. For the period-\( 3 \) sequence, however, the
diagonal fixed point is completely unstable, and the two-cycle should
be responsible for the critical behavior, as in the case of the spin
models \cite{haddad3}. In this case, the scaling analysis must be
somewhat adapted to take into account that two renormalization-group
iterations are needed in order that the system goes back to the vicinity
of one of the two points that are part of the two-cycle \cite{derrida}.
The result is simply that the specific-heat critical exponent is now
given by\begin{equation}
\label{alpha-ciclo}
\alpha =2-2\frac{\ln p}{\ln \Lambda },
\end{equation}
where \( \Lambda  \) is the leading eigenvalue of the linearized
second-iterate of the RG recursion relations about any one of the
two points of the attractor. Results of this analysis have already
been given elsewhere \cite{haddad1}, and will not be repeated here.

For the model with Rudin-Shapiro aperiodic interactions, we remarked
above that there exist two non-diagonal fixed points together with
the curve composed of two-cycles, and that the linearization of the
second iterates of the recursion relations about any point on the
two-cycles curve gives the same eigenvalues. For each \( q>2+\sqrt{2} \),
we may therefore determine numerically which value of \( \alpha  \)
Eq. (\ref{alpha-ciclo}) predicts, and then compare it with the direct
analysis of the singularity which comes from the TM method. It is
also possible to obtain \( \alpha  \) in the usual way, linearizing
the recursion relations (first iterate) about the non-diagonal fixed
points, using the leading eigenvalue that comes from the solution
of Eq. (\ref{eq2.18}). The coincidence of the values is already an
important hint of the correctness of scaling predictions, and we have
indeed verified it for several choices of \( q \).

\begin{figure}
\begin{center}
{\centering \resizebox*{7cm}{7cm}{\includegraphics{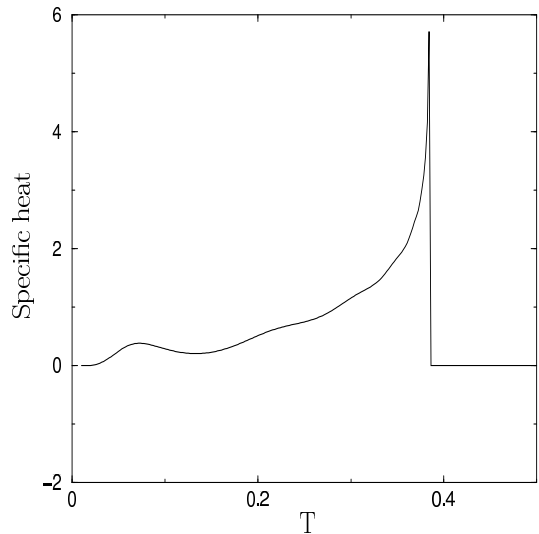}} \par}
\end{center}

\caption{Specific heat of the model with Rudin-Shapiro aperiodicity, in a lattice 
with $p=2$ and $q=4$, calculated using the transfer-matrix technique. 
In this figure $\epsilon_d$ is 100 times greater than the other energies.}
\end{figure}

In Figure 6 we show the TM results for the specific heat in a lattice
with \( q=4 \), with a certain choice of interaction energies. The
first interesting feature is the appearance of log-periodic oscillations
in the low-temperature phase, as in spin models \cite{andrade1}. This is a
natural consequence of the discrete scale-invariance of the aperiodic sequence
(due to its self-similar character), which implies a natural rescaling factor 
in the renormalization group \cite{sornette}.
Several different values of the interaction energies must be separately
analysed, and the net result is an exponent \( \alpha =0.252\pm 0.08 \).
Two points must be carefully stressed: first, that some choices for
the energies (\( \epsilon _{a}=\epsilon _{d} \) and \( \epsilon _{b}=\epsilon _{c} \),
for instance) give rise to an effectively periodic model, because
of the symmetries of the Rudin-Shapiro rule, and should therefore
be kept out of the analysis; second, \( \alpha  \) does not show
any important dependence on the values (provided they are not of the
form that makes the model effectively periodic, of course), which
points to a true {}``aperiodic universality class'' associated with
the Rudin-Shapiro geometrical perturbation of the model.Now, for \( q=4 \), 
the scaling result is \( \alpha =0.253692\ldots  \), in striking agreement 
with the TM value. The same scenario is present for several other values of \( q \)
we have tested, what leads us to believe in the correctness of the methods. 

In conclusion, we have presented detailed renormalization-group and transfer-matrix 
calculations  for a class of interacting polymer models on diamond-like hierarchical 
lattices, with  aperiodically distributed coupling constants. Although straightforward, 
the exact renormalization-group analysis has revealed a surprising family of attractors 
in case of Rudin-Shapiro aperiodicity, and this prompted us to resort to the 
transfer-matrix formalism to check the scaling results. However, we had to develop a 
complete reformulation of this method, in order to be able to apply it to the polymer 
problem. The transfer-matrix calculations have confirmed the results of the simple 
scaling analysis, but have also revealed peculiarities of the transition that were 
not accessible to the renormalization-group study. What is most important is to
notice that aperiodic perturbations may lead to new universality classes, adding
up to the usual criteria of dimensionality and symmetry. In a sense, the breakdown
of translation invariance may be relevant to the determination of new universal
behaviors, and the particular way in which this invariance is broken
must be taken into account. The introduction of disorder, for example, is a way
of breaking translation invariance, and there are several instances in which its
effects on critical behavior are well-known. Aperiodic distributions of couplings
are just another way of accomplishing this, and, although more difficult to
implement physically, they are amenable to more controlled calculations, such 
as those presented in this paper.

\ack{We thank Andr\'{e} P. Vieira and Suani T. R. Pinho for useful discussions.
This work has been supported by Brazilian agencies Fapesp and CNPq.}

\end{document}